\documentclass[aps,prd,twocolumn,tightenlines,showpacs,aps,amsmath,amssymb,nofootinbib]{revtex4}

\usepackage{latexsym}
\usepackage{graphicx}
\usepackage{subfigure}
\usepackage{cases}

\usepackage{hyperref}
\usepackage{amssymb}
\usepackage{bm}
\usepackage{natbib}

\newcommand{\beq}{\begin{equation}}
\newcommand{\eeq}{\end{equation}}
\newcommand{\beqa}{\begin{eqnarray}}
\newcommand{\eeqa}{\end{eqnarray}}
\newcommand{\xs}{x_{\rm s}}
\newcommand{\xres}{x_{\rm res}}

\begin{document}
\title{Constraints on Superconducting Cosmic Strings from Early Reionization}

\author{Hiroyuki Tashiro}
\email{Hiroyuki.Tashiro@asu.edu}
\author{Eray Sabancilar}
\email{Eray.Sabancilar@asu.edu}
\author{Tanmay Vachaspati}
\email{tvachasp@asu.edu}

\affiliation{
Physics Department, Arizona State University, Tempe, Arizona 85287, USA.
}


\begin{abstract}
Electromagnetic radiation from superconducting cosmic string loops can
reionize neutral hydrogen in the universe at very early epochs, and
affect the cosmic microwave background (CMB) temperature and
polarization correlation functions at large angular scales. We
constrain the string tension and current using WMAP7 data, and compare
with earlier constraints that employed CMB spectral distortions. Over
a wide range of string tensions, the current on the string has to be
less than $\sim 10^7~{\rm GeV}$.
\end{abstract}


\maketitle

\section{Introduction}

Spontaneous breaking of symmetries in unified theories of particle 
physics can lead to the formation of cosmic strings (for a review 
see, e.g., \cite{LindeBook,VilenkinBook,Copeland09,Copeland11}). In
certain models, the cosmic strings can be superconducting
\cite{Witten:1984eb}.
%

As superconducting cosmic strings pass through magnetized regions, 
they accumulate electric currents, and emit electromagnetic 
radiation \cite{Vilenkin:1986zz} and particles
\cite{Witten:1984eb,Berezinsky:2009xf}.
Therefore, the presence 
of currents on strings leads to a number of observational signatures 
that can be probed
by present and future experiments.
Radiation from strings
is not isotropic but is most efficient at cusps --- parts of a 
string that move very close to the speed of light --- and kinks 
--- discontinuities in the tangent vector to the string. The
strongly beamed radiation has been studied as sources of various 
observable effects such as gamma ray 
bursts \cite{Paczynski,Berezinsky:2001cp}, ultra
high energy neutrinos \cite{Berezinsky:2009xf}, radio transients
\cite{Vachaspati:2008su,Cai:2011bi}, and distortions of the cosmic
microwave background (CMB) \cite{Tashiro:2012nb,Sanchez89,Sanchez90}.
In addition to these signatures that are particular to superconducting
strings, there are more generic effects that are gravitational and
do not rely on superconductivity. These include the anisotropy of the 
CMB temperature \cite{Dvorkin11} and B-mode plarization \cite{Pogosian08,Avgoustidis:2011ax},
gravitational
waves \cite{Damour01,Damour05,pulsar,Olmez,Battye,Dufaux12}, gravitational 
lensing \cite{lensing}, wakes in the 21cm maps 
\cite{Brandenberger:2010hn,Hernandez:2011ym,McDonough:2011er,Hernandez:2012qs,Pagano:2012cx}, early structure formation \cite{Shlaer:2012rj}, early reionization due to early structure
formation \cite{Olum:2006at}, and formation of dark matter clumps \cite{Berezinsky:2011wf}.

In this paper, we focus on the reionization of cosmic hydrogen gas in 
the intergalactic medium (IGM) due to photons emitted from superconducting
cosmic string loops. Hydrogen gas in the IGM was electrically
neutral after cosmic recombination, which occurs at a redshift 
$z_{\rm rec}\approx 1100$. However, as luminous structures, such as stars 
and galaxies form, ultra-violet (UV) 
photons emitted by these objects ionize surrounding hydrogen gas.
Finally, hydrogen gas in the IGM is totally ``reionized''
\cite{Barkana:2000fd}. The observation of the Gunn-Peterson effect
\cite{1965ApJ...142.1633G} in high redshift
quasar spectra, which probes the amount of neutral hydrogen along
the line of sight, suggests
that full reionization was accomplished by $z \sim 6$ \cite{Fan:2005es}.
Recently, WMAP measured the CMB anisotropy, and obtained the optical
depth due to Thomson scattering, $\tau\sim 0.088$, in WMAP 7-year
data \cite{Komatsu:2010fb}. This result implies that the cosmological
reionization process began around $z=10$, at least. 

%
%

In contrast to UV emission due to structure formation, superconducting 
cosmic string loops emit ionizing photons even at high redshifts, $z>100$,
when there are a few luminous objects which can ionize hydrogen. Therefore, 
the existence of superconducting cosmic strings can cause early reionization. 
Once early reionization occurs, it leaves footprints on the CMB temperature 
and polarization anisotropy spectra. By using Markov Chain Monte Carlo 
(MCMC) analysis, and taking into account the effect of superconducting 
cosmic string loops, we evaluate the CMB anisotropy spectra, and compare 
them with WMAP 7-year data. 

This paper is organized as follows. We calculate the radiation emitted by 
superconducting cosmic string loops in Sec.~\ref{sec:uvphoton}, and in
Sec.~\ref{sec:Hion} we find the ionization fraction due to UV photons 
emitted by cosmic string loops. In Sec.~\ref{sec:Hion}, we also calculate 
the optical depth for Thomson scattering of CMB photons on free electrons, 
and compare the ionization history with strings to the standard history 
without strings used by WMAP analysis. 
In Sec.~\ref{sec:anisotropy}, we estimate the effects of string loops on 
CMB anisotropies. In conclusion, we obtain constraints on the string tension, 
$G\mu$, and the electric current, $I$, on the string in 
Sec.~\ref{conclusions}. 

Throughout this paper, we use parameters for a flat $\Lambda$CDM model: 
$h=0.7$ $(H_0=h \times 100 ~ {\rm km /s /Mpc})$, $\Omega_b=0.05$ and 
$\Omega_m=0.26$. Note also that $1+z=\sqrt{t^{\prime}/t}$ in the radiation 
dominated epoch, and $1+z=(1+z_{\rm eq}) (t_{\rm eq}/t)^{2/3}$ in the 
matter dominant epoch, where $t^{\prime} = (2 \sqrt{\Omega_r} H_0)^{-1}$ 
and $z_{\rm eq}=\Omega_m / \Omega_r$ with 
$h^2 \Omega_r=4.18 \times 10^{-5}$. We also adopt natural units, 
$\hbar = c =1$.


\section{Reionization from Cosmic String Loops}
\label{sec:reion}


\subsection{UV Photon Emission from Loops}
\label{sec:uvphoton}

Solutions to the dynamical equations for cosmic strings show that a cosmic
string loop often develops cusps and almost always has kinks. These features
lead to the emission of high energy photons from superconducting cosmic
strings. The number of photons emitted from a cusp per unit time per unit 
frequency is
\cite{Cai:2011bi}
\beq\label{fc}
f_{\omega}^{c} \equiv \frac{d^{2} N_{c}}{d \omega dt} \sim \frac{4 \pi}{3} 
                  \frac{I^{2} L^{1/3}}{\omega^{5/3}},
\eeq
and from a kink \cite{radiokink}
\beq\label{fk}
f_{\omega}^{k} \equiv \frac{d^{2} N_{k}}{d \omega dt} 
               \sim 2 \pi \psi \frac{I^{2}} {\omega^2},
\eeq
where $I$ is the current, $L$ is the loop length, $\psi$ is the sharpness of the kink, and $\omega$ is the 
frequency of the emitted photon. 

Note that the radiated electromagnetic power from a single cusp 
dominates over the power radiated from a single kink by a factor 
$\sim (\omega L)^{1/3} \gg 1$, since the loop length is much larger
than the radiation wavelength of interest. 

The spectrum of photons emitted from cosmic string loops per unit physical 
volume is
\beq
\frac{d^2 \mathcal{N}}{d \omega dt} = \int  dn(L,t) \frac{d^{2} N}{d \omega dt},
\eeq
where $dn(L,t)$ is the number density of loops in the matter dominated era 
\begin{equation}
dn(L,t) = \frac{\kappa C_L dL}{t^2 (L+\Gamma G\mu t)^2}\ , 
              \ \ t > t_{\rm eq},
\label{eq:ndensity}
\end{equation}
where $\kappa \sim 1$, and 
$C_L \equiv  1 + \sqrt{t_{\rm eq}/(L+\Gamma G\mu t)}$. 
The second term in $C_L$ accounts for loops in the matter dominated
era that are left over from the radiation dominated era. We note that
the distribution of cosmic string loops is not completely established.
Although analytical studies
\cite{Rocha:2007ni,Polchinski07,Dubath08,Vanchurin11,Lorenz:2010sm} and cosmic string
network simulations \cite{Bennett90,Allen90,Hindmarsh97,Martins06,Ringeval07,
Vanchurin06,Olum07,Shlaer10,Shlaer11} agree on the distribution of large 
loops, there is disagreement on the distribution of small loops. 
We have checked that a different choice of loop distribution, {\it e.g.}, 
as suggested in Ref.~\cite{Lorenz:2010sm}, does not change our final 
constraints significantly.

Then, the number of photons emitted from all
cusps per unit time per unit volume per unit frequency is given by
\begin{equation}
 {d^2 \mathcal{N}_{c} \over d \omega dt}
  \approx {8 \pi \kappa {\cal C} \over 9 }  
\left( t_{eq} \over t \right )^{1/2}  {I^2 \over (\Gamma G \mu)^{7/6} t^{8/3}} 
      \, \omega^{-5/3},
 \label{eq:photon_spectrum} 
\end{equation}
in the matter dominated epoch, where $\Gamma$ is determined by the dominant
energy loss mechanism, as discussed below Eq.~(\ref{P_gamma}), 
and ${\cal C}$ is the average number of cusps per loop. 
The number of kinks per loop is expected
to be quite large, and the radiation will also depend on the sharpness of 
the kinks $\psi$. The effective number of kinks per loop, i.e., weighting a 
kink by its sharpness, will be denoted by ${\cal K}$, and then,
the number of photons emitted per unit time per unit volume per unit
frequency from kinks is
\beq
 \frac{d^2 \mathcal{N}_{k}} {d \omega dt} \approx 
 \frac{4 \pi \kappa {\cal K}}{3}  \left(t_{eq} \over t \right )^{1/2} \frac{I^2}{(\Gamma G \mu)^{3/2} t^{3}} 
         \, \omega^{-2}.
\eeq
Unless ${\cal K}$ is larger than ${\cal C} (\omega \Gamma G\mu t)^{1/3}$,
the number of photons emitted by cusps is larger than the number emitted
by kinks. We shall assume that cusps dominate the emission from string
loops. 

Cosmic string loops also emit gravitational waves with power
\cite{VilenkinBook}
\beq
P_{g} \sim \Gamma_{g} G \mu^{2},
\eeq
where $\Gamma_{g} \sim 50$, whereas the power due to electromagnetic radiation
from superconducting loop cusps is \cite{VilenkinBook}
\beq\label{P_gamma}
P_{\gamma} \sim \Gamma_{\gamma} I \sqrt{\mu},
\eeq
where $\Gamma_{\gamma} \sim 10$. $P_{g} \sim P_{\gamma}$ at the critical value
of the current
\beq
I_{*} \equiv \frac{\Gamma_{g} G \mu^{3/2}}{\Gamma_{\gamma}}.
\eeq
The lifetime of a loop is determined by the dominant energy loss mechanism, and
can be estimated as
\beq
\tau = \frac{\mu L}{P} \equiv \frac{L}{\Gamma G \mu},
\eeq
where
\begin{numcases}
{\Gamma \approx}
\Gamma_g  & 
\qquad for $I < I_*$
\label{larger}
\\ 
\Gamma_g \frac{I}{I_*}  & \qquad for $I > I_*$.
\label{smaller}
\end{numcases}


\subsection{Ionization Fraction}
\label{sec:Hion}

%

UV photons emitted from a loop of cosmic string ionize hydrogen in the
surrounding medium since the mean free path of UV photons is very short.
If we assume that one emitted ionizing photon ionizes one
hydrogen atom, then the ionization rate in the region surrounding a
loop is equal to the number of ionizing photons emitted from the loop 
per unit time
\begin{equation}
{d N_{i} \over dt} = \int _ {\omega_i} ^{\omega_c}  d\omega~ f_\omega,
\label{eq:def-uv}
\end{equation}
where $f_\omega$ is the spectrum of the photons emitted from a loop of
cosmic string given by Eq.~(\ref{fc}), and the integration is from the
Lyman $\alpha$ frequency, $\omega_i=13.6~$eV, to a high frequency
cutoff, $\omega_c$. Photons with energy higher than $\omega_c$
do not contribute to the photo--ionization, because the cross-section
decreases rapidly as the photon energy increases \cite{Zdziarski89}. 
Since the number of ionizing photons falls off with frequency, our 
results are not very sensitive to the exact value of $\omega_c$, and
we will set $\omega_c=10^4~$eV.

The recombination rate in the volume surrounding the loop is 
\begin{equation}
{d n_r \over dt}  = n_e n_{\rm HII} \alpha_r  = n_{\rm H}^2 \alpha_r ,
\label{eq:dnrdt}
\end{equation}
where $\alpha_r$ is the recombination coefficient. Typically 
$\alpha_r=2.6 \times 10^{-13}\, {\rm cm}^3/ {\rm s}$ at $T_e=10^4~$K, and 
$n_{\rm H}=8.42 \times 10^{-6}\, (1+z)^3\, \Omega_B h^2~ {\rm cm}^{-3} $. 
In order to get the final expression in Eq.~(\ref{eq:dnrdt}),
we assume that all the hydrogen inside the surrounding volume
is ionized, i.e., $n_{\rm HII}=n_e = n_{\rm{H}}$.

The steady-state assumption between the ionization and recombination rates 
in the surrounding region allows us to determine the ionized volume
around a loop of length $L$,
\begin{equation}
V(L) = {d N_{i} /dt  \over d n_r/dt},
\end{equation}
where the numerator is given in Eq.~(\ref{eq:def-uv}), and the denominator
in Eq.~(\ref{eq:dnrdt}). This is the volume around a single loop and depends
on the length, $L$, of the loop. We can obtain the ionization fraction
due to all cosmic string loops, $\xs$, by multiplying the number density 
of loops, $dn(L,t)$, by $V(L)$, and integrating over $L$,
\beqa
 \xs &=& \int dn(L,t) V(L) \nonumber \\
     &=&  {1 \over  n_{\rm H} ^2 \alpha} 
      \int _ {\omega_i} ^{\omega_c}  d\omega \int dn(L,t) f_\omega(L,t) 
                          \nonumber \\ 
     &\equiv& {1 \over  n_{\rm H} ^2 \alpha} 
\int _ {\omega_i} ^{\omega_c}  d\omega \frac{d^2 \mathcal{N}_c}{d \omega dt}. 
\label{eq:xi_eq}
\eeqa  
Using Eq.~(\ref{eq:photon_spectrum}), we can perform the integration
to get 
\begin{equation}
 \xs (z) = {4 \pi \kappa \over 3 \alpha n_H ^2 (z) }
\left(t_{eq}
  \over t \right )^{1/2} {{\cal S} \over t^{8/3}} 
\left(\omega_{i} ^{-2/3} - \omega_{c} ^{-2/3} \right), 
\label{eq:xi_redshift}
\end{equation}
where $z \sim z_{eq} (t/t_{eq})^{-2/3}$ in the matter dominated epoch and
\begin{equation}
{\cal S} = \frac{I^2}{(\Gamma G \mu)^{7/6}}.
\label{eq:Sdefn}
\end{equation}
Note that the reionization signature from strings only depends on the
particular combination of string tension and electric current that occurs
in ${\cal S}$. Hence, observations will impose constraints on ${\cal S}$.
In contrast to the standard scenario where structure formation 
suddenly results in reionization, cosmic strings cause early reionization
that builds up gradually. From Eq.~(\ref{eq:xi_redshift}), we see that
the ionization due to cosmic strings grows with decreasing redshift
as
\begin{equation}\label{xsmat}
\xs \propto z^{-5/4}\ , \ \ z \gg 1 \ .
\end{equation}

\begin{figure}[b]
\includegraphics[scale=0.5]{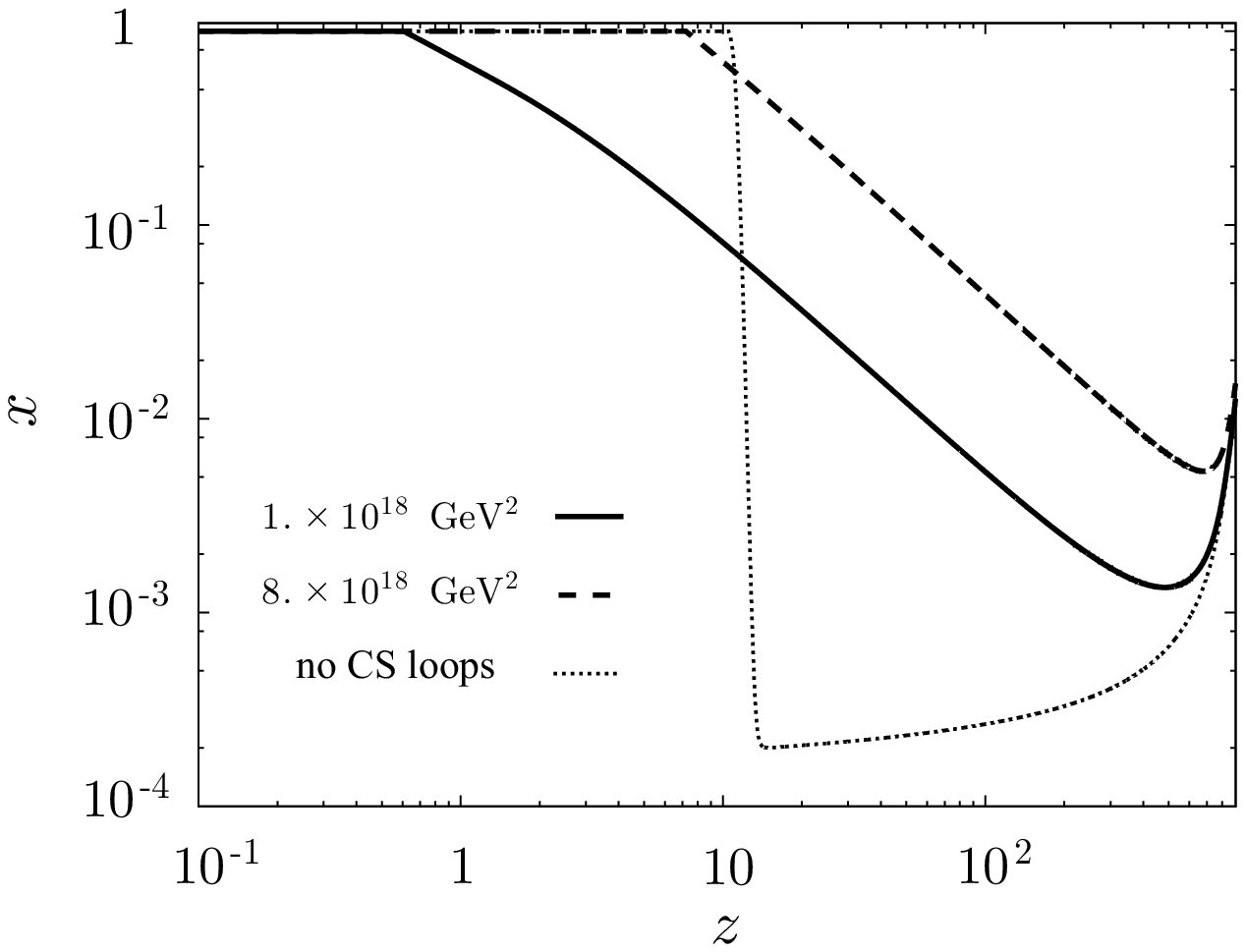} 
\includegraphics[scale=0.5]{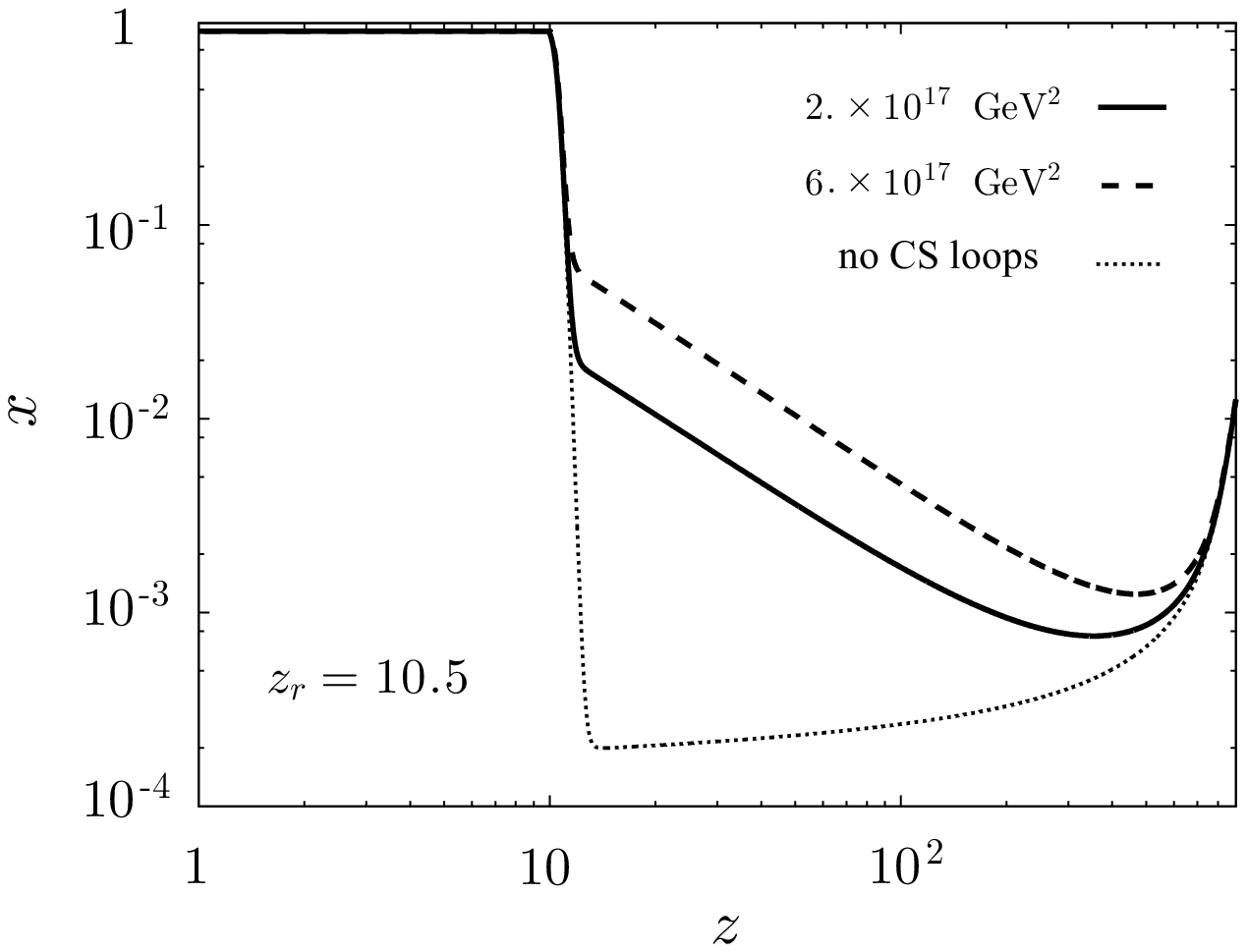} 
 \caption{The dotted line shows the reionization fraction in the absence 
of cosmic strings, as modeled by WMAP. The solid and dashed curves in
the top panel show the reionization fraction from cosmic strings, $\xs(z)$, 
for two different values of the parameter ${\cal S}$ defined in 
Eq.~(\ref{eq:Sdefn}).
In the bottom panel we plot the total reionization fraction, $x(z)$, 
which is a sum of the structure formation, residual, and string components
for a few different string parameters.
}
\label{fig:csl_ionization}
\end{figure}
\begin{figure}[t]
  \includegraphics[scale=0.5]{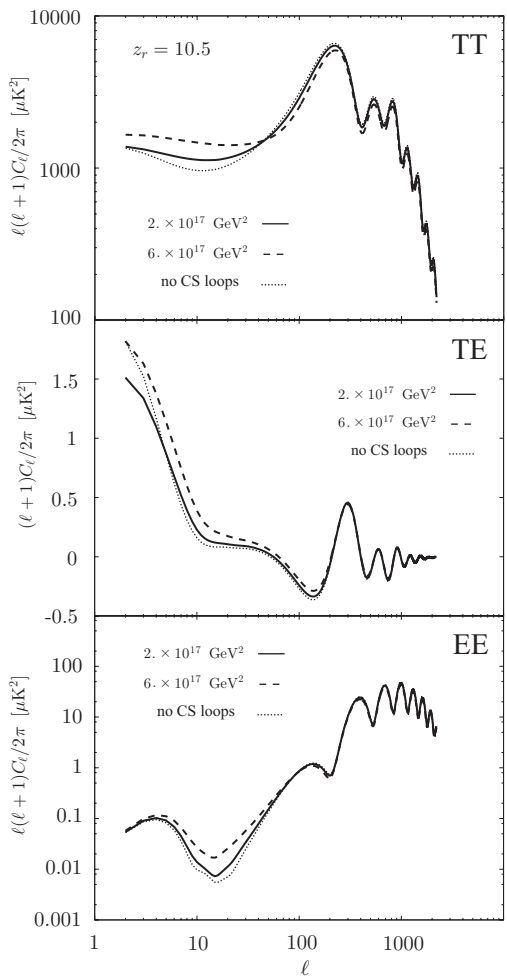}
 \caption{CMB anisotropy spectra with fixed $z_{r}=10.5$.
The panels from top to bottom 
show $TT$, $TE$ and $EE$ angular spectra,
respectively. The solid line is for cosmic string loops with 
${\cal S}=2  \times 10^{17} ~{\rm GeV}^2$, for which the optical depth 
is $\tau =0.1$. The dashed line represents the case for 
${\cal S} =6 \times 10^{17} ~{\rm GeV}^2$, for which the optical depth is $0.13$. 
For comparison, we plot the WMAP recommended reionization model (dotted curve)
with the $\tanh$-shape reionization with $z_{r}=10.5$.
}
\label{fig:zreCMB}
\end{figure}
\begin{figure}[t]
\includegraphics[scale=0.5]{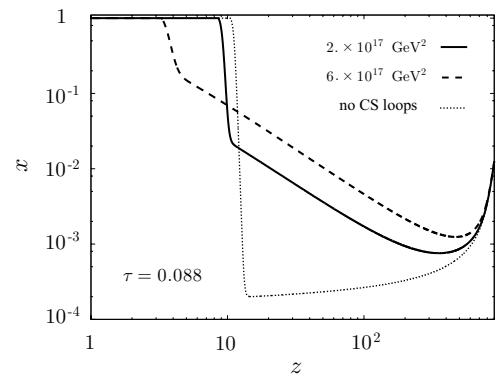} 
 \caption{Possible reionization histories with fixed optical depth $\tau=0.09$,
without strings (dotted) and with strings for two values of ${\cal S}$.
}
\label{xztaufixed}
\end{figure}

The reionization due to strings is in addition to the reionization due
to structure formation. Hence, the total ionization fraction is
\begin{equation}
x = (x_0+\xres ) + \xs,
\label{xtotal}
\end{equation}
where $x_0$ denotes the ionization fraction due to structure formation,
and $\xres$ the residual ionization from the epoch of recombination.
The reionization fractions are functions of redshift, and the WMAP7 team 
has modeled $x_0(z)$ as
\begin{equation}
x_0(z) = \frac{1}{2} \left [
           1+{\rm tanh}\left ( \frac{2 (1+z_r)^{3/2} -(1+z)^{3/2} }{3\Delta
			z  (1+z_r)^{1/2}} \right )
                      \right ],
\label{wmapmodel}
\end{equation}
with $z_r \sim 10$ and $\Delta z \sim 0.5$. Note that the total
ionization fraction cannot exceed unity. Our result in 
Eq.~(\ref{eq:xi_redshift}) however, can exceed 1 because we 
have assumed that the volumes reionized by different loops do not 
overlap. This assumption becomes invalid at high reionization fraction, and so 
$x$ is cutoff at 1 if Eq.~(\ref{xtotal}) evaluates to a value larger 
than 1. In Fig.~\ref{fig:csl_ionization}, we show 
$x_0+\xres$ and $\xs$, and also the total reionization fraction, $x$,
as functions of $z$.


We can also calculate the optical depth for Thomson scattering as
\begin{equation}
 \tau = \int_{t_{rec}}  ^{t_0} dt ~n_e \sigma_T,
\end{equation}
where $t_{rec}$ is the time of recombination epoch, $t_0$ is the present age of
the universe, $\sigma_T$ is Thomson scattering cross-section. The optical depth
for the solid line in the top panel of Fig.~\ref{fig:csl_ionization} reaches
$\tau \sim 0.09$, which is consistent with WMAP recommended value. However, the
reionization by cosmic string loops is very slow. For comparison, we plot a
$\tanh$-step ionization fraction parameterized by the middle point $z_r$ and
duration $\Delta z$  as used by the WMAP 7-year analysis \cite{Komatsu:2010fb}.
This toy model in Fig.~\ref{fig:csl_ionization} gives the optical depth $\tau
\sim 0.09$. 

In the matter dominated epoch, the evolution of the ionization fraction due to 
cosmic string loops is given by Eq.~(\ref{xsmat}). As a result, even at 
$z \sim 2$, the ionization fraction is less than 0.5 for 
${\cal{S}}=10^{18}~ {\rm GeV}^2$. This result conflicts with Gunn-Peterson 
test for high redshift quasars that 
the ionization fraction at $z\sim 6$ is larger than 0.9 \cite{Fan:2005es}.

To satisfy the Gunn-Peterson test, at least 
${\cal S} = 8 \times 10^{18} ~{\rm GeV}^2$ is
required as shown with the dashed line in the top panel of
Fig.~\ref{fig:csl_ionization}. However, the optical depth is
$\tau \sim 0.7$ in this case. Thus, it is difficult to fit the CMB anisotropy
spectrum in this model to the WMAP data, where we use cosmological 
parameters that are consistent with other cosmological observations. 
Therefore, we can conclude that the electromagnetic radiation from 
cosmic string loops is not a primary source for the cosmological reionization. A more consistent model with cosmic string loops, which
satisfies the constraints of both the
optical depth and Gunn-Peterson test, is the tanh-shape model with
subdominant contribution of cosmic string loops.
This leads to the reionization
histories shown in the bottom panel of Fig.~\ref{fig:csl_ionization}.

Before closing this section, we comment on the assumptions we made when
evaluating the ionized volume $V$. We have ignored the velocities of cosmic 
string loops, and then evaluated the ionized volume $V$, assuming that a 
cosmic string loop ionizes the same region continuously. Since a cosmic 
string loop is expected to have relativistic center of mass velocity, 
the ionized region will continually shift with the motion of the cosmic 
string loop. However, we are interested in the average ionized fraction
and the location of the ionized volume at any instant of time does
not matter. Thus, we can neglect the velocity of cosmic string loops.
Note that as cosmic strings move they also create wakes \cite{Shlaer:2012rj}, 
and can lead to early structure formation \cite{Shlaer:2012rj}, which can
indirecly affect the reionization history \cite{Olum:2006at}.

\begin{figure}[h]
  \includegraphics[scale=0.5]{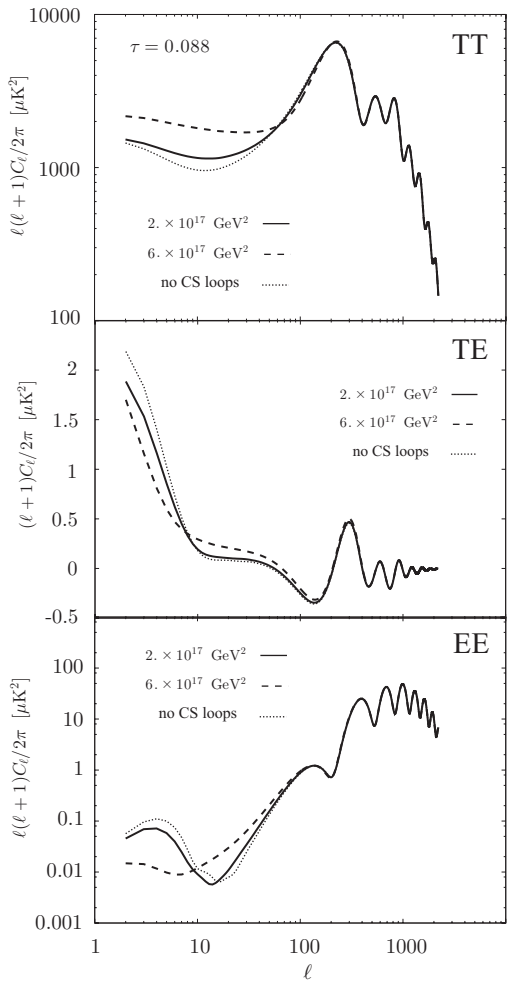}
 \caption{The CMB E-mode polarization anisotropy spectra with fixed
 optical depth $\tau=0.088$.
The solid line is for cosmic string loops with ${\cal S}=2
 \times 10^{17} ~{\rm GeV}^2$ and $z_{r}=9.1$.
The dashed line is for ${\cal S} =6 \times 10^{17}~{\rm GeV}^2$
and $z_{r}=3.5$.
For comparison, we plot the case without cosmic string loops as the dotted line.}
\label{fig:optCMB}
\end{figure}

In the integration in Eq.~(\ref{eq:def-uv}), we have ignored photons with 
energy higher than $\omega_c = 10^4~$eV since the universe is transparent 
to photons with higher energies. However, cosmological redshift can 
decrease the energies of such photons, and reduce them to below $\omega_c$. 
Even if we take this redshift effect into account, the optical depth of photo
ionization is negligibly-small \cite{Zdziarski89}. Therefore, we can ignore
photons with energy higher than $10^4~$eV since they do not contribute to the
ionization process.


\section{Effect on CMB anisotropy}
\label{sec:anisotropy}


Reionization of the cosmic medium affects the CMB temperature and polarization
correlators. Thus, CMB observations can lead to constraints on superconducting
cosmic strings. We calculate CMB anisotropy with modified ${\texttt{CAMB}}$ code
\cite{Lewis:1999bs} to take into account the effect of cosmic string loops on
cosmic reionization. 

We first fix the parameter $z_r=10.5$ in Eq.~(\ref{wmapmodel}). As discussed 
above and shown in Fig.~\ref{fig:csl_ionization},
this leads to different reionization histories with different optical depths.
The results for the CMB correlators are shown in Fig.~\ref{fig:zreCMB}. The
high optical depth due to cosmic string loops suppresses the amplitude of the
acoustic oscillations in the temperature anisotropy as shown in the top panel
of Fig.~\ref{fig:zreCMB}. 
The high optical depth enhances the amplitude
on small $\ell$ in both the temperature and polarization anisotropy spectra,
because Thomson scattering in the induced early reionization generate
additional CMB anisotropy.

Next, we fix the optical depth to $\tau = 0.088$, and let $z_r$ be an
independent parameter. Since the optical depth is an integral over
the ionization fraction, it depends on both $z_r$ and ${\cal S}$, and
fixing $\tau$ leads to a relation between $z_r$ and ${\cal S}$. For
example, $z_{r}=9.1$ for  ${\cal S}=2 \times 10^{17}$, and $z_{r}=3.5$ for
${\cal S}=6 \times 10^{17}$. On the other hand, $\tau = 0.088$ 
requires $z_{r} \sim 11$ without cosmic string loops. The redshift 
evolution of the ionization fraction for this case is
shown in Fig.~\ref{xztaufixed}.

\begin{figure}[t]
\includegraphics[scale=0.5]{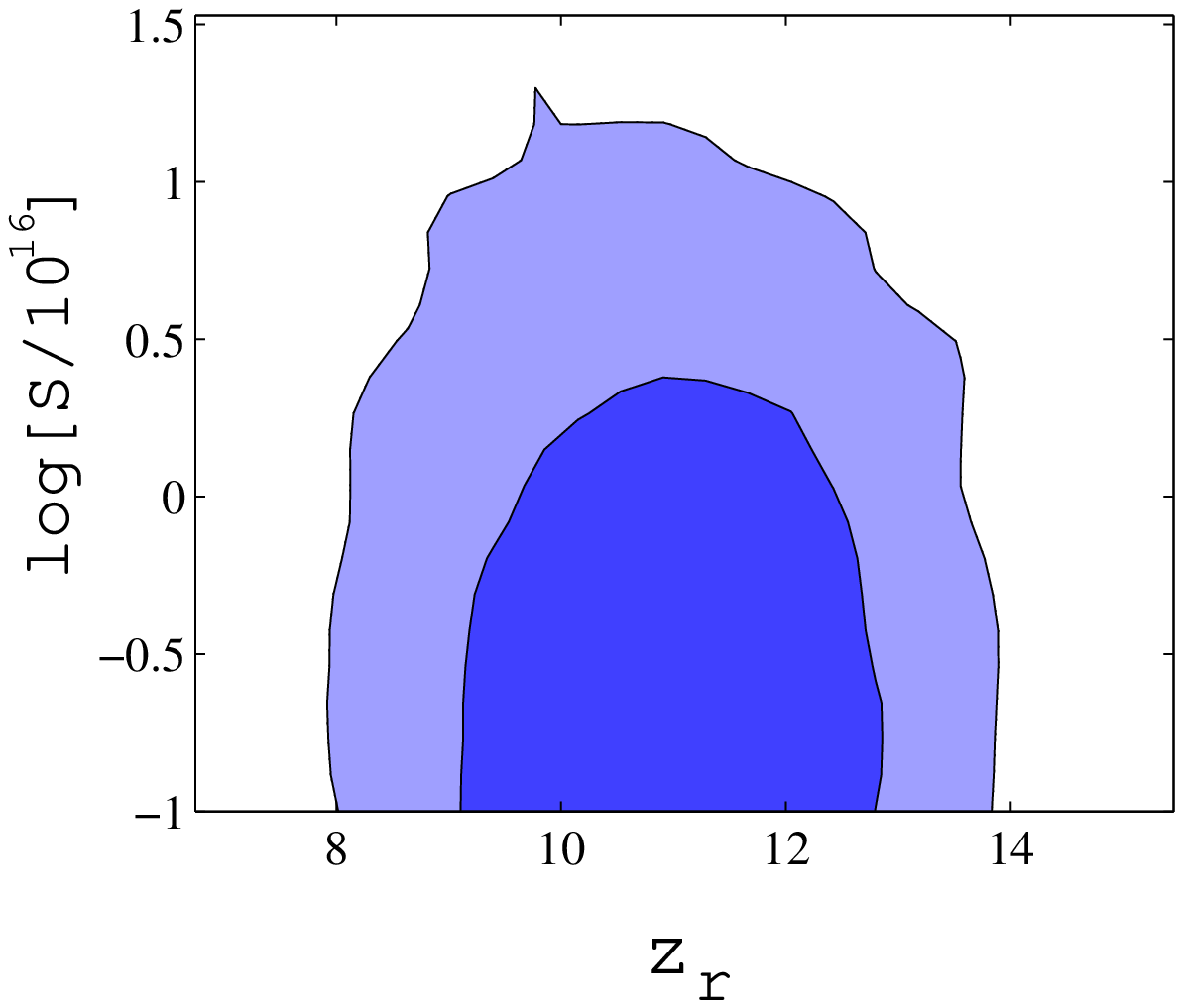}
\includegraphics[scale=0.5]{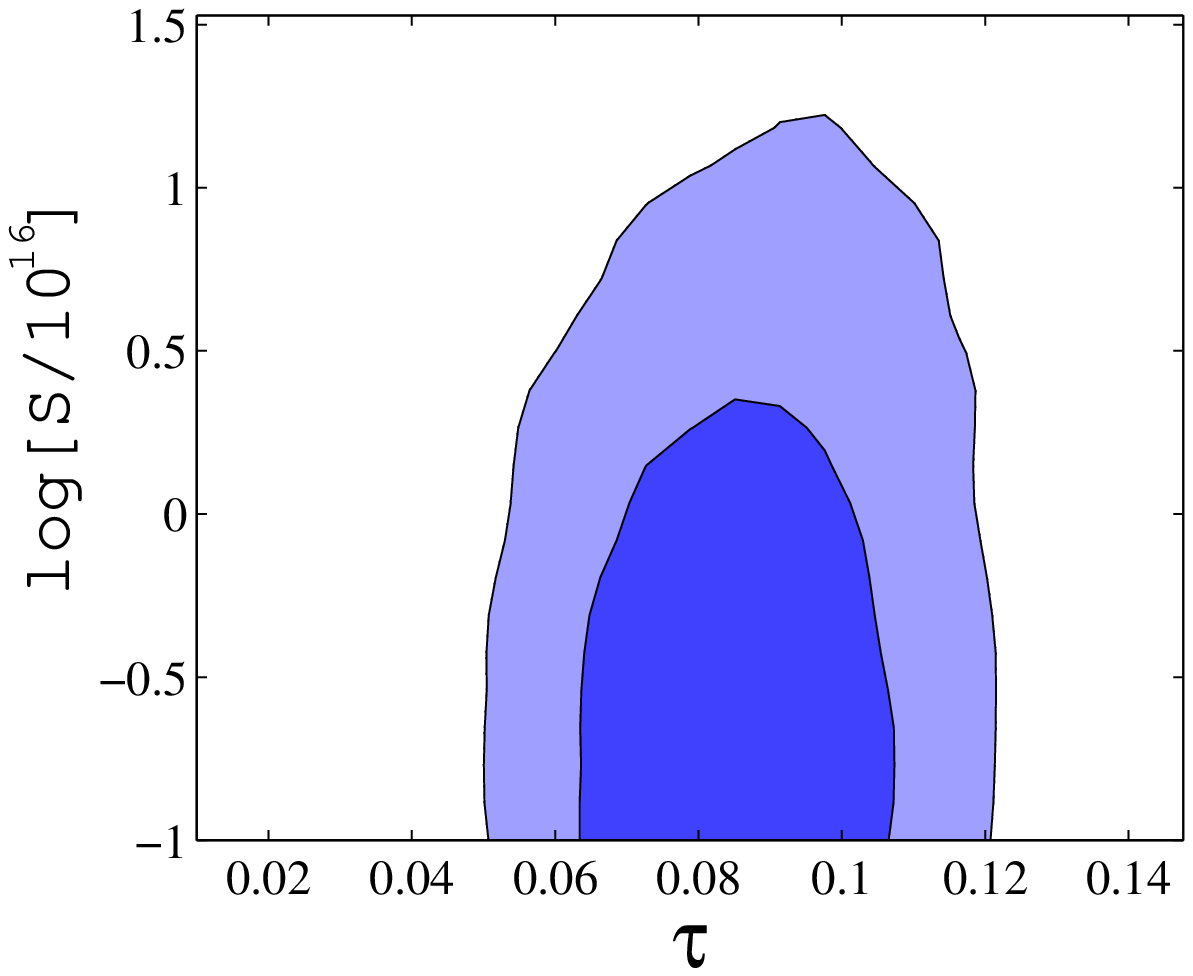}
\caption{ The 2-D contour of marginalized probabilities, in the ${\cal S}-z_r$ 
plane on the top, and the ${\cal S}-\tau$ plane on the bottom. The contours from the
inside to the outside show the 1-sigma and 2-sigma confidence level.}
\label{fig:2-dcontour}
\end{figure}

The CMB anisotropy spectra are shown in Fig.~\ref{fig:optCMB}. Larger
values of ${\cal S}$ enhance the amplitude of CMB temperature anisotropy
at large angular scales (small $\ell$). The peak position of $E$-mode 
polarization on small $\ell$ depends on the scale of the quadrupole 
component of CMB temperature anisotropy at the reionization epoch, 
$z_r$. Earlier reionization, corresponding to larger $\cal S$, implies 
that this peak occurs at lower $\ell$. 

\section{Conclusions}
\label{conclusions}
\begin{figure}[t]
 \includegraphics[scale=0.54]{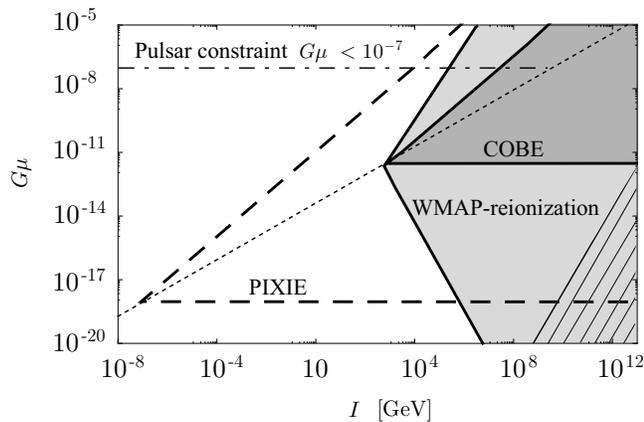} 
\caption{
 The reionization constraint on the $I$--$G \mu$ plane shown in conjunction with
the CMB spectral distortion and other constraints. The lightly shaded
region is ruled out from WMAP 7-year result. CMB $\mu$-distortion constraints
by COBE-FIRAS measurement rule out the dark shaded area. The millisecond 
pulsar observations constrain $G\mu \lesssim 10^{-7}$ and is shown as
the dot-dashed line. The region above the diagonal thin dashed line is 
where gravitational radiation dominates over electromagnetic radiation, 
{\it i.e.}, $I < I_*$. Possible future constraints from CMB $\mu$-distortion 
are also shown by the dashed line if no such effect is seen by PIXIE.
The hatched region is excluded because 
$I > \sqrt{\mu_s}$, and exceeds the saturation value of the current on superconducting strings.
} 
\label{fig:string_const}
\end{figure}

In order to get the constraint on cosmic string loops from WMAP 7 year data, we
take MCMC analysis in the $\Lambda$CDM flat universe model with modified
{\texttt{COSMOMC}} code \cite{Lewis:2002ah} to take into account the
reionization effect of cosmic string loops. Fig.~\ref{fig:2-dcontour} shows 2-D
contour of marginalized probabilities. Recall that $z_{r}$ is the
redshift at which $\tanh$-shape reionization fraction is 0.5. From WMAP 7
year data, we obtain the constraint ${\cal S} < 3 \times 10^{16}~{\rm GeV}^2$
at the 1-$\sigma$ confidence level.

Accordingly, from Eq.~(\ref{eq:Sdefn}), we obtain the constraint 
\begin{equation}
 G \mu \gtrsim 10^{-11 } \left( { 10^3~ {\rm GeV} \over I}\right) ^{7/12},
\end{equation}
for the gravitational radiation dominant case, $I<I_*$, and
\begin{equation}
 G \mu \lesssim 4 \times 10^{-13 },
\end{equation}
for the electromagnetic radiation dominant case, $I>I_*$. 

We plot these constraints in the $I$--$G\mu$ plane in Fig.~\ref{fig:string_const} 
as the lightly shaded region. For comparison we
also show the constraint from $\mu$-distortion with COBE-FIRAS
result as the dark shaded region, and possible future constraint from PIXIE where the interior region of the dashed triangle can be ruled out if no spectral distortion is observed by PIXIE, which was studied in Ref.~\cite{Tashiro:2012nb}.  

Fig.~\ref{fig:string_const} shows that reionization constraints 
on superconducting strings are much tighter than the constraints
from CMB spectral distortions. In future, the PIXIE experiment to
measure CMB spectral distortions at very low levels will be able
to impose constraints that are stronger than current reionization
constraints. Even then, reionization constraints will be stronger
for light strings with large currents.
 
\acknowledgments
This work was supported by the DOE and by NSF grant PHY-0854827 at 
ASU. We are also grateful for computing resources at the ASU Advanced 
Computing Center (A2C2).

\bibstyle{aps}

\end{document}